\documentclass[prl,twocolumn,superscriptaddress,longbibliography]{revtex4-1}
\usepackage{amsmath,amssymb}
\usepackage{mathtools}
\usepackage{simplewick}

\usepackage[applemac]{inputenc}
\usepackage[english]{babel}
\usepackage[pdftex]{graphicx}
\usepackage{color}
\usepackage{dcolumn}
\usepackage{bm}
\usepackage{mathbbol}
\usepackage{array,multirow,graphicx}
\usepackage{amsfonts}
\usepackage{times}
\usepackage{xcolor}

\begin{document}
\title{Quantum and Classical Lyapunov Exponents in Atom-Field Interaction Systems}
\author{Jorge Ch\'avez-Carlos}
\affiliation{Instituto de Ciencias Nucleares, Universidad Nacional Aut\'onoma de M\'exico, Apdo. Postal 70-543, C.P. 04510  Cd. Mx., M\'exico}
\author{B. L\'opez-del-Carpio} 
\affiliation{Instituto de Ciencias Nucleares, Universidad Nacional Aut\'onoma de M\'exico, Apdo. Postal 70-543, C.P. 04510  Cd. Mx., M\'exico}
\author{Miguel A. Bastarrachea-Magnani}
\affiliation{Physikalisches Institut, Albert-Ludwigs-Universitat Freiburg, Hermann-Herder-Str. 3, Freiburg, Germany, D-79104.} 
\author{Pavel Str\'ansk\'y}
\affiliation{Faculty of Mathematics and Physics, Charles University, V Hole\v{s}ovi\v{c}k\'ach 2, Prague 180 00, Czech Republic}
\author{Sergio Lerma-Hern\'andez}
\affiliation{Facultad de F\'{i}sica, Universidad Veracruzana, Circuito Aguirre Beltr\'an s/n, Xalapa, Veracruz  91000,  Mexico}
\author{Lea F. Santos}
\affiliation{Department of Physics, Yeshiva University, New York, New York 10016, USA.} 
\author{Jorge G. Hirsch} 
\affiliation{Instituto de Ciencias Nucleares, Universidad Nacional Aut\'onoma de M\'exico, Apdo. Postal 70-543, C.P. 04510  Cd. Mx., M\'exico}
%%%%%%%%%%%%%%%%%%%%%%%%%%%%%%%%%%%%%%%%

\begin{abstract}
The exponential growth of the out-of-time-ordered correlator (OTOC)  has been proposed as a quantum signature of classical chaos. The growth rate is expected to coincide with the classical Lyapunov exponent. This quantum-classical correspondence has been corroborated for the kicked rotor and the stadium billiard, which are one-body chaotic systems. The conjecture has not yet been validated for realistic systems with interactions. We make progress in this direction by studying the OTOC in the Dicke model, where two-level atoms cooperatively interact with a quantized radiation field. For parameters where the model is chaotic in the classical limit, the OTOC increases exponentially in time with a rate that closely follows  the classical Lyapunov exponent.
\end{abstract} 

\maketitle

%%%%%%%%%%%%%%%% INTRODUCTION %%%%%%%%%%%%%%%%%

Quantum chaos tries to bridge quantum and classical mechanics. The search for quantum signatures of classical chaos has ranged from level statistics~\cite{HaakeBook,StockmannBook} and the structure of the eigenstates~\cite{Chirikov1985,Flambaum1999b} to the exponential increase of complexity~\cite{Peres1996,BorgonoviARXIV} and the exponential decay of the overlap of two wave packets~\cite{Jalabert2001,Cucchietti2002,Gorin2006,Fine2014}. Recently, the pursuit of exponential instabilities in the quantum domain has been revived by the conjecture of a bound on the rate growth of the out-of-time-ordered correlator (OTOC) \cite{Maldacena2016PRD,Maldacena2016JHEP}. First introduced in the context of superconductivity~\cite{Larkin1969}, the OTOC is now presented as a measure of quantum chaos, with its growth rate being associated with the classical Lyapunov exponent. The OTOC is not only a theoretical quantity, but has also been measured experimentally via nuclear magnetic resonance techniques~\cite{Garttner2017,Li2017,Wei2018,NiknamARXIV}.

The correspondence between the OTOC growth rate and the classical Lyapunov exponent has been explicitly shown in two cases of one-body chaotic systems, the kicked-rotor~\cite{Rozenbaum2017} and, after a first unsuccessful attempt~\cite{Hashimoto2017}, the stadium billiard~\cite{RozenbaumARXIV}. It was also achieved for chaotic maps~\cite{GarciaARXIV}. For interacting many-body systems, while exponential behaviors for the OTOC have been found for the Sachdev-Ye-Kitaev model~\cite{Maldacena2016PRD,Bagrets2017} and for the Bose-Hubbard model~\cite{Bohrdt2017,Shen2017}, a direct demonstration of the quantum-classical correspondence has not yet been made. Studies in this direction include~\cite{Elsayed2015,Akila2017,Bianchi2018,ScaffidiARXIV,BorgonoviARXIV,TarkhovARXIV} and~\cite{Rammensee2018}.

Here, we investigate the OTOC for the Dicke model~\cite{Dicke1954,Garraway2011}. Comparing with one-body systems, the model is a step up toward an explicit quantum-classical correspondence for interacting many-body systems, since it contains $N$ atoms interacting with a quantized field. 

The Dicke model was originally proposed to explain the collective phenomenon of superradiance: the field mediates interatomic interactions, which causes the atoms to act collectively~\cite{Dicke1954,Hepp1973,*Wang1973,*Carmichael1973}. Superradiance has been experimentally studied with ultracold atoms in optical cavities~\cite{Baumann2010,Baumann2011,Ritsch2013,Baden2014,Klinder2015,Kollar2017}. The model has also found applications beyond superradiance in various different fields. It has been employed, for instance, in studies of ground-state and excited-state quantum phase transitions~\cite{Hepp1973,*Wang1973,*Carmichael1973,Castanos2005,Fernandez2011b,Brandes2013,Bastarrachea2014a,Larson2017}, entanglement creation~\cite{Schneider2002,*Lambert2004,*Kloc2017}, nonequilibrium dynamics~\cite{Fernandez2011,Altland2012PRL,Lerma2018,Kloc2018}, quantum chaos~\cite{Lewenkopf1991,Emary2003PRL,*Emary2003,Bastarrachea2014b,*Bastarrachea2015,*Bastarrachea2016PRE,Chavez2016}, and monodromy~\cite{Babelon2009,Kloc2017JPA}. Recently, the model has received revived attention due to new experiments with ion traps~\cite{Cohn2018,SafaviARXIV} and the analysis of the OTOC~\cite{AlaviradARXIV,LewisARXIV}.

In the classical limit, the Dicke model presents regular and chaotic regions depending on the Hamiltonian parameters and excitation energies~\cite{Chavez2016}. This allows us to benchmark the OTOC growth against the presence and absence of chaos. The results in the chaotic region display three different temporal behaviors: a sinusoidal evolution at short times, followed by an exponential growth, that holds up to the saturation of the dynamics.  Our approach, based on the use of an efficient basis for the convergence of the eigenstates, enables the treatment of systems that are large enough to reveal the exponential part of the dynamics. We find that the exponential growth rate is in close agreement with the classical Lyapunov exponent. 

{\em Quantum and Classical Hamiltonian.--} The Dicke model has $N$ two-level atoms of level spacing $\omega_0$ coupled with a single mode of a quantized radiation field of frequency $\omega$. The Hamiltonian is given by
\begin{equation}
\hat{H}_{D}=\dfrac{\omega}{2}(\hat{p}^2+\hat{q}^2)+\omega_{0}\hat{J}_{z}+2 \frac{\gamma}{\sqrt{N/2}} \hat{J}_{x} \, \hat{q}- \frac{\omega}{2},
\label{eq:HD}
\end{equation}
where $\hbar=1$; $\hat{q}=(\hat{a}^{\dagger} + \hat{a})/\sqrt{2}$ and
$\hat{p}=i(\hat{a}^{\dagger} - \hat{a})/\sqrt{2}$ are the quadratures of the bosonic field and $\hat{a} (\hat{a}^{\dagger})$ is the annihilation (creation) operator;  $\gamma$ is the atom-field interaction strength; and the collective atomic pseudo-spin operators, $\hat{J}_{x,y,z} =(1/2)\sum_{n=1}^N \sigma_{x,y,z}^{(n)} $, are the sums of the Pauli matrices for each atom $n$. The eigenvalue of the total spin operator $\mathbf{\hat{J}}^2 =  \hat{J}_x^2 + \hat{J}_y^2 + \hat{J}_z^2$ is $j(j+1)$. The critical point $\gamma_{c} = \sqrt{N/(2j)} \sqrt{\omega\omega_{0}}/2$ marks the transition from a normal phase ($\gamma<\gamma_{c}$) to a superradiant phase ($\gamma>\gamma_{c}$). We set $\omega=\omega_0=1$ in the illustrations below and work with the symmetric atomic subspace ($j=N/2$), where the ground state lies. The model has two degrees of freedom.

The classical Hamiltonian is built by employing Bloch coherent states and Glauber coherent states~\cite{Ribeiro2006,Bakemeier2013,Chavez2016}. The first are given by $ |z\rangle=\left(1+\left|z\right|^{2}\right)^{-j} e^{z \hat{J}_+}|j, -j\rangle ,$ where $z\in \mathbb{C}$ and $|j, -j\rangle$ is the ground state for the atoms. The Glauber coherent states are $|\alpha\rangle=e^{-|\alpha|^2/2}e^{\alpha \hat{a}^\dagger}|0\rangle ,$ where  $\alpha\in \mathbb{C}$ and $|0\rangle$ is the photon vacuum. The canonical variables $(p,q)$ and $(j_z,\phi)$ are given in terms of the coherent state parameters $\alpha=\sqrt{\frac{j}{2}}(q+ip)$ and $ z=\sqrt{\frac{1+j_z}{1-j_z}}e^{-i \phi}$, respectively. Deriving the classical Hamiltonian is basically equivalent to replacing the operators with the canonical variables $(q,p)$ and $(j_z,\phi)$ as $\hat{q}\rightarrow \sqrt{j} q$, $\hat{p}\rightarrow \sqrt{j} p$, $\hat{J}_{z}\rightarrow j j_z$, $\hat{J}_{x}\rightarrow j\sqrt{1-j_{z}^{2}}\,\cos\phi$. It reads
\begin{equation}
\label{Eq:hcl}
H_{D}^{cl} =  j \frac{\omega}{2}\left(p^{2}+q^{2}\right)+j \omega_{0}\,j_{z}+2 j \gamma \sqrt{1-j_{z}^{2}}\,q\,\cos\phi.
\end{equation}
Since the classical limit is reached for $j\rightarrow \infty$,  the  effective Planck constant is $\hbar_{eff}=1/j$.

We denote the energy per particle as $\epsilon = H_{D}^{cl}/j$, which is independent of $j$. Since the number of bosons in the field is unlimited, the range of values of $\epsilon$ is only limited from below. The ground state energy 
is given by $\epsilon_0(\gamma) = -\omega_0$ for $\gamma\leq \gamma_c$ and by $\epsilon_0 (\gamma) = -\frac{\omega_0 }{2} \left(\frac{\gamma_c^2}{\gamma^2}+ \frac{\gamma^2}{\gamma_c^2}\right) $ for $\gamma > \gamma_c $. 
 
\begin{figure}[ht] 
\centering
\includegraphics[width=0.45\textwidth]{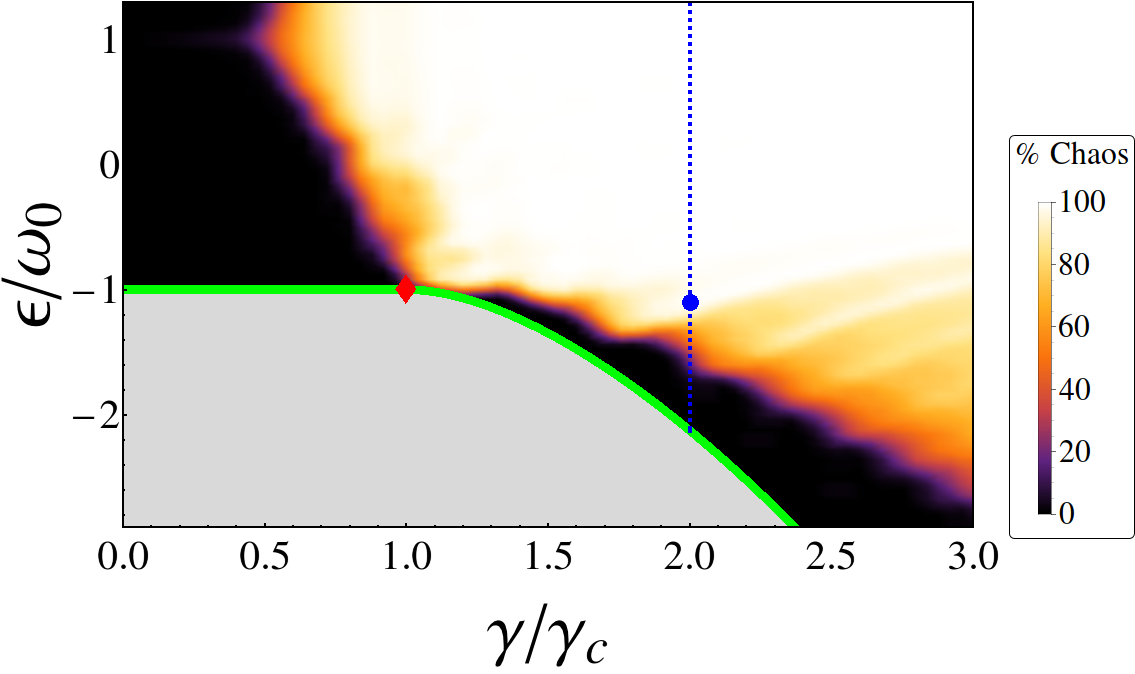} 
\caption{ 
Percentage of chaos over energy shells as a function of energy and coupling strength. The thick (green) solid line follows the ground state energy and the diamond marks the critical point. The (blue) vertical dotted line indicates the coupling $\gamma=2\gamma_c$ and the circle marks the energy chosen for the studies below.  
}
\label{fig:chaos}
\end{figure}

With the classical Hamiltonian, we obtain a map of the degree of chaoticity of the system as a function of the energy $\epsilon$ and the  interaction strength $\gamma$, as shown in Fig.~\ref{fig:chaos}. The task of drawing the map is quite demanding. For each value of $\epsilon$ and $\gamma$, we consider a large sample of initial conditions distributed homogeneously in the energy shell. The Lyapunov exponent $\lambda_{cl}$ is evaluated for each initial condition solving the dynamical equations and the fundamental matrix simultaneously~\cite{Chavez2016}. If  $\lambda_{cl}>0$, the initial condition is chaotic and for $\lambda_{cl}=0$, the initial condition is regular. The percentage of chaos is defined as the ratio of the number of chaotic initial conditions over the total number of initial conditions in the sample.  This percentage is shown in Fig.~\ref{fig:chaos} with a color gradient: dark indicates that most initial conditions are regular  and light indicates that most are chaotic. (Notice that one should look only at the results above the thick solid line that marks the ground state.) Regularity predominates for $\gamma/\gamma_c< 0.6$. For $\gamma/\gamma_c> 0.6$, most regular trajectories have low energies, while large energies are associated with chaos. This map guides our analysis of the OTOC below.

{\em Method.--}
The OTOC quantifies the degree of non-commutativity in time between two Hermitian operators with small or null commutator at time $t = 0$. In terms of position and momentum, it is written as
\begin{equation}
C_n^{qp} (t)= -\langle \Psi_n | \left[q(t),p(0) \right]^2 | \Psi_n \rangle ,
 \label{Eq:OTOC}
\end{equation} 
where $|\Psi_n \rangle$ and $E_n$ are the eigenstates and eigenvalues of $\hat{H}_D$.  In Ref.~\cite{Hashimoto2017},  $C_n^{qp} (t)$ is called microcanonical OTOC.  We refer to the exponential growth rate of the OTOC as $\Lambda_{\text{Q}}$. In the semiclassical limit, substituting the commutator by the Poisson bracket, one gets for a classically chaotic system, $\{q(t),p(0) \} =\partial q(t)/\partial q(0) \sim e^{\lambda_{cl} t}$, where $\lambda_{cl}$ is the classical Lyapunov exponent. This suggests the connection between $\Lambda_{\text{Q}}$ and $\lambda_{cl}$, and justifies referring to  $\Lambda_{\text{Q}}$ as the quantum Lyapunov exponent.

Using the temporal evolution of the operator $\hat{q}(t)=e^{iHt}\hat{q}e^{-iHt}$, Eq.~(\ref{Eq:OTOC}) can be expressed as~\cite{Hashimoto2017}
\begin{equation}
 C^{qp}_n (t)=\sum_l b_{nl}(t)b^*_{nl}(t) ,
\label{eq:motocb}
\end{equation}
where the matrix elements 
\begin{eqnarray}
b_{nl}(t) &=& -i\langle \Psi_n|\left[\hat{q}(t),\hat{p}(0) \right]|\Psi_l\rangle \nonumber \\
             &=& -i\sum_k (e^{i\Omega_{nk}t}q_{nk}p_{kl}-e^{i\Omega_{kl}t}p_{nk}q_{kl}) , \nonumber
\end{eqnarray}             
with $q_{nk}=\langle \Psi_n|\hat{q}|\Psi_k\rangle$, $p_{nk}=\langle \Psi_n|\hat{p}|\Psi_k\rangle$, and $\Omega_{nk}=E_n-E_k$.
Since the Dicke Hamiltonian is of the form  $\hat{H}_D= \omega \hat{p}^2/2+ V(\hat{q})$ and $[\hat{H}_D, \hat{q}] = - i \omega \hat{p}$,  
\begin{equation}
 b_{nl}(t)=\dfrac{1}{\omega}\sum_k q_{nk}q_{kl}(\Omega_{kl}e^{i\Omega_{nk}t}-\Omega_{nk}e^{i\Omega_{kl}t}),
 \label{eq:bn}
\end{equation}
which simplifies the calculations. The OTOC is obtained by evaluating numerically only the matrix elements of $\hat{q}$ in the energy eigenbasis. For this, instead of employing the usual photon number (Fock) basis,  we resort to an efficient basis that guarantees convergence of the eigenvalues and wave functions for a broad part of the spectrum (see~\cite{Bastarrachea2014PSa,*Bastarrachea2014PSb}).   

{\em Quantum Lyapunov Exponent.--}  In this Letter, we concentrate our analysis on chaotic eigenstates. They are chosen along the vertical line in Fig.~\ref{fig:chaos}, where the coupling parameter is strong, $\gamma= 2\gamma_c$. This line exhibits regular and chaotic regions. From the ground state $\epsilon_0=-2.125$ to $\epsilon\approx -1.6$, the dynamics is regular. From $\epsilon\approx -1.6$ to $\epsilon\approx -1.2$, regular and chaotic trajectories coexist. For larger energies, $\epsilon>-1.2$, chaos cover almost the whole energy shell. We select a group of fifty-one eigenstates in the chaotic energy region with $E_n /(j \omega_0) \in (-1.11,-1.09)$. They are indicated with a circle in Fig.~\ref{fig:chaos}.

\begin{figure}[ht]
\centering
\begin{tabular}{c} 
\includegraphics[width=0.4\textwidth]{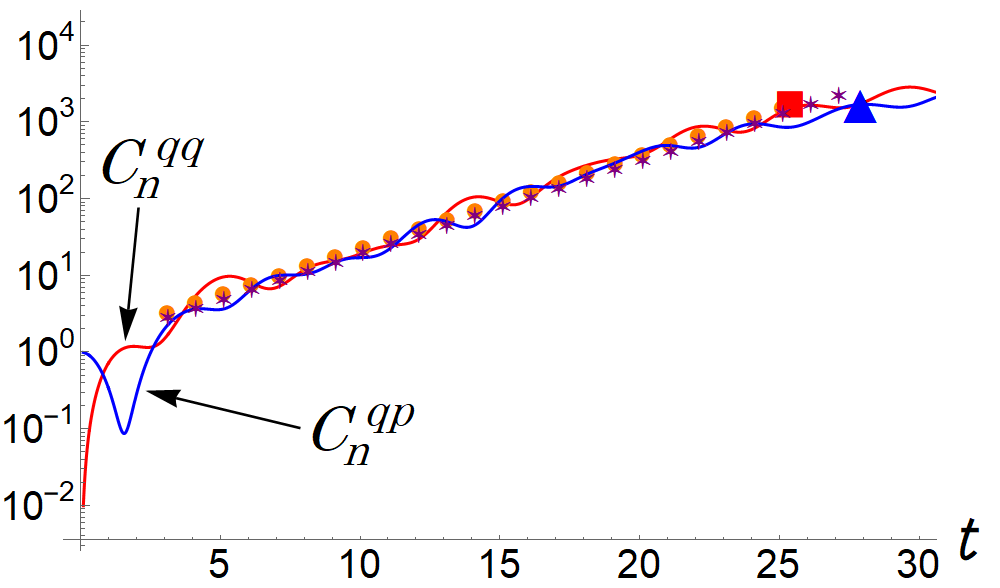} \\ 
\large (a) \\
\includegraphics[width=0.4\textwidth]{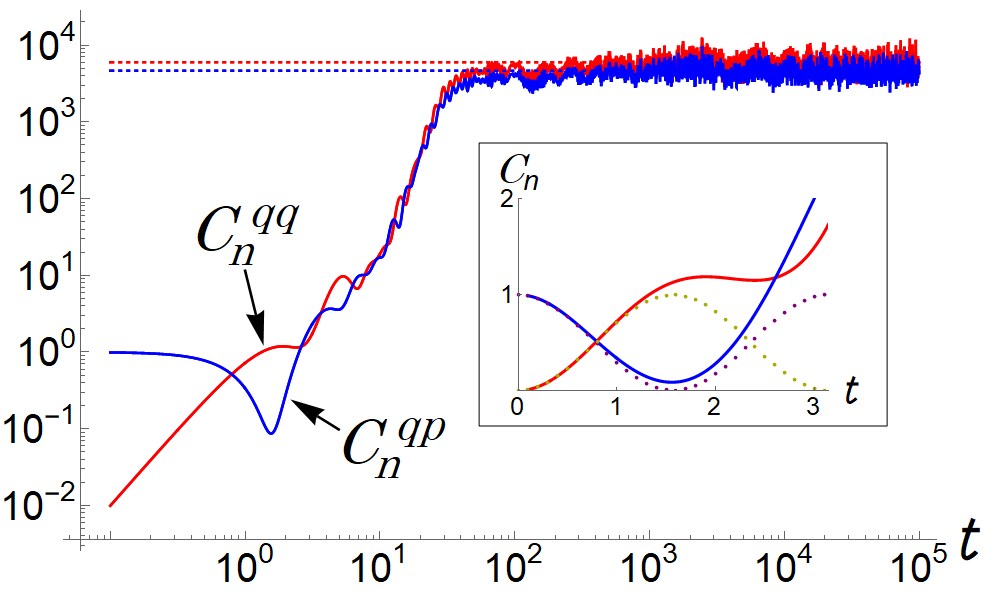} \\ 
\large (b) \normalsize \\
\end{tabular}
\caption{Panel (a): Exponential growth of the OTOC  for an eigenstate with $E_n/(j \omega_0) \approx -1.1$; numerical results (solid line), fit for $C^{qp}_n (t)$ (stars)  and for $C^{qq}_n (t)$ (circles); saturation times (square and triangle). 
Panel (b): Log-log plot for the evolution of the OTOC and saturation value (dotted lines). Inset: short time behavior compared with $\sin^2(t)$ and $\cos^2(t)$ (dotted lines). We use $j=100$, $n=1625$.}
\label{fig:OTOC}
\end{figure}

In Fig.~\ref{fig:OTOC}~(a), we show that  even for a single representative eigenstate, the behavior of the OTOC  is clearly exponential from $t \gtrsim \pi/\omega_0$ up to the saturation of the dynamics. The  growth rate $\Lambda_{\text{Q}}= 0.139$ is obtained by fitting the curve with a straight line indicated with stars in the figure. 

The exponential behavior is robust with respect to two different probes:

(i) It holds when we use the commutator for the operator $\hat{q}$ at different times, $C^{qq}_n (t)= -\langle \Psi_n | \left[q(t),q(0) \right]^2 | \Psi_n \rangle $, as also shown in Fig.~\ref{fig:OTOC}~(a). The associated fit, indicated with circles, provides $\Lambda_{\text{Q}}' = 0.139$. Both exponential fits lead, within the numerical uncertainty, to the same quantum Lyapunov exponents. 

(ii) The exponential growth rates are very similar for the fifty-one different states selected in the chaotic region.

The log-log plot in Fig.~\ref{fig:OTOC}~(b) makes evident the appearance of different behaviors at different time scales. For $t < \pi/\omega_0$, the dynamics of $C^{qp}_n (t)$ [similarly for $C^{qq}_n (t)$] is controlled by the diagonal matrix elements in Eq.~(\ref{eq:bn}), $b_{nn}(t)= (2 /\omega)  \sum_k  q_{kn}^2 \Omega_{kn}  \cos(\Omega_{kn}t)$, with few states contributing significantly to the sum, all with energy differences $\Omega \approx 1.0$. The short-time evolution is therefore approximately described by the square of a cosine function [sine for $C^{qq}_n (t)$]. The two sinusoidal curves are shown with dotted lines in the inset of Fig.~\ref{fig:OTOC}~(b).

At long times, the quantum dynamics saturates to the infinite-time average,
\begin{equation}
\overline{C_n^{pq}}=\dfrac{1}{\omega^2}\sum_{k,l} q^2_{nk}q^2_{kl}\left(\Omega^2_{kl}+ \Omega^2_{nk}\right),
\end{equation}
which is obtained from Eqs.~(\ref{Eq:OTOC}) and (\ref{eq:bn}) using that $\overline{\exp[i (\Omega_{ij} - \Omega_{kl})t} =0 $ for $\Omega_{ij} \neq \Omega_{kl}$. $\overline{C_n^{pq}}$ and $\overline{C_n^{qq}}$ are shown in Fig. \ref{fig:OTOC}~(b) with dotted lines. These averages are related with the square of the size of the available phase space~\cite{Hashimoto2017}. For the Dicke Hamiltonian, it scales with $j^2$ and with the number of bosons in the system, which grows with the excitation energy.

After the exponential growth, the OTOC fluctuates around its asymptotic value, as seen in Fig.~\ref{fig:OTOC}~(b), with a standard deviation $\sigma$. We define the saturation time $t_S$ as the time when the OTOC reaches for the first time the value  $\overline{C_n^{pq}} - \sigma $. The values of $t_S$ for $C^{qp}_n (t) $ and $C^{qq}_n (t) $ are marked in Fig.~\ref{fig:OTOC}~(a) with a triangle and a square, respectively. The saturation time marks the point beyond which quantum effects are strong and the quantum-classical correspondence no longer holds, therefore the association between $t_S$ and the Ehrenfest time. The saturation of the dynamics for finite quantum systems is in contrast to what one finds for classical systems, where the spectrum is continuous. As $j$ increases and the system approaches the classical limit, $\overline{C_n^{pq}}$ grows and $t_S$  increases with it. 

\begin{figure}[h]
\includegraphics[width=0.4\textwidth]{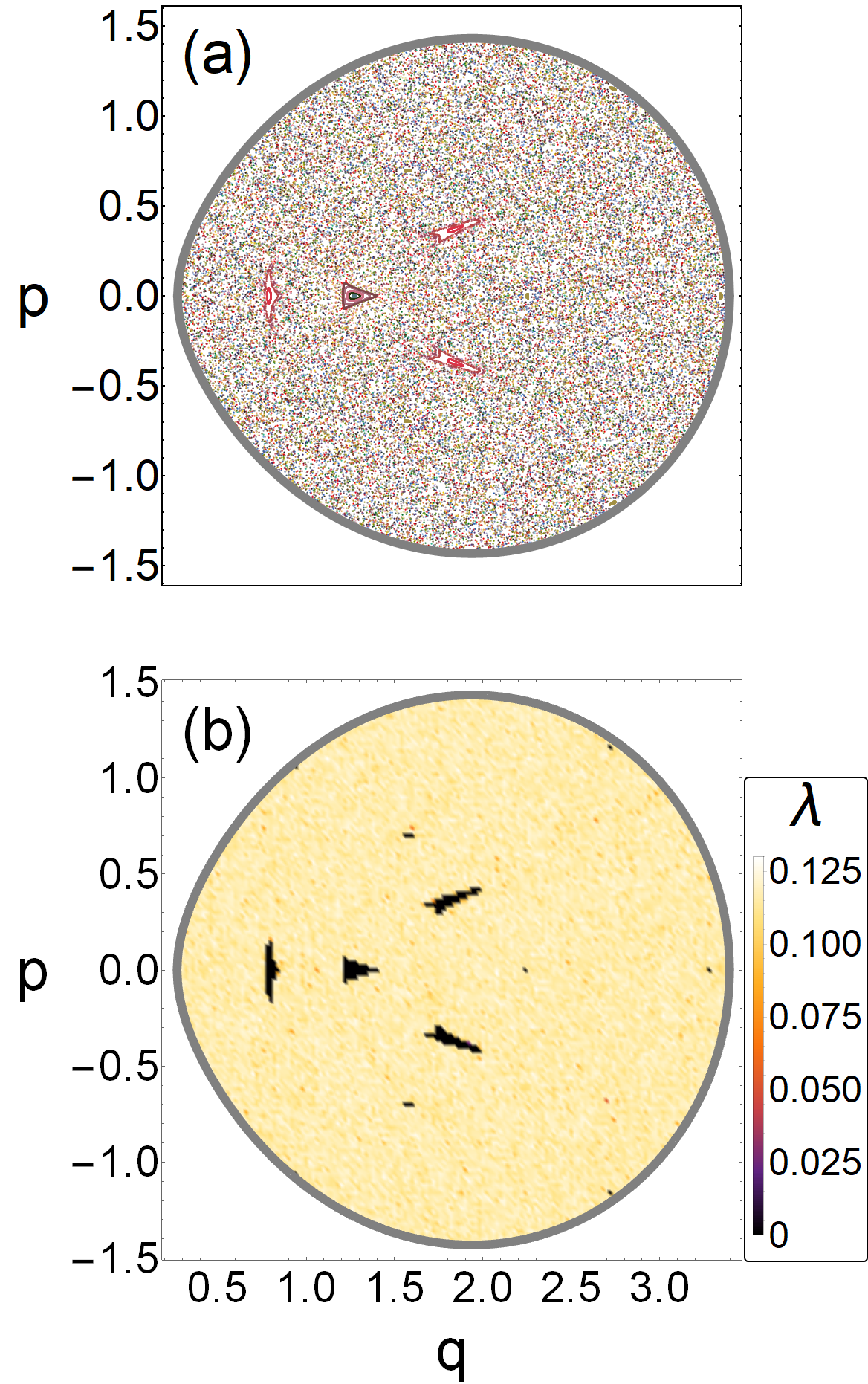} \\ 
\caption{Panel (a): Poincar\'e surfaces of section projected on the plane $(q,p)$ for $\phi=0$ and energy $-1.1 \omega_0$ for various initial conditions. Panel (b): map of chaos over the same Poincar\'e surface in terms of the finite time classical Lyapunov exponents. These exponents are evaluated for each trajectory up to $10\,000$ units of time, which is enough to have stable results. }
\label{fig:Hus_ns}
\end{figure}

{\em Quantum-classical correspondence.--} In a fully chaotic system, there is one classical Lyapunov exponent associated with the whole energy shell. Numerically, however, the Lyapunov exponents are computed for finite times, so they depend on the initial conditions.  We evaluated the time average of the exponents for each trajectory up to $10\,000$ units of time, which is enough to have stable results. The trajectories for several initial conditions are depicted in Fig.~\ref{fig:Hus_ns}~(a). This figure shows the  Poincar\'e surfaces of section projected on the plane $(q,p)$ for $\phi=0$ and energy $-1.1 \omega_0$. In addition to chaotic trajectories, one identifies also regular trajectories. These islands of stability are clearly visible in Fig.~\ref{fig:Hus_ns}~(b) as small black regions. This bottom panel is a classical map of chaos for the same energy and plane of Fig.~\ref{fig:Hus_ns}~(a). The color code represents the values of the finite time Lyapunov exponents obtained for each initial point in the phase space. 

We consider thousands of initial conditions, from which a large number $N_{ch}$ is chaotic. To obtain a single value for the Lyapunov exponent for the chaotic region of the energy shell, we average over those initial conditions that give rise to classical chaotic trajectories and discard those with zero exponents. We then have,
\begin{equation}
\tilde{\lambda}_{\langle \ln \rangle}=  \frac{1}{N_{ch}}\sum_{k=1}^{N_{ch}}
\lambda_k = \lim_{t\rightarrow\infty} \frac{1}{t}  \frac{1}{N_{ch}}\sum_{k=1}^{N_{ch}} 
 \ln (e^{\lambda_k t}).
\label{Eq:aveLog}
\end{equation}
The purpose of writing the last term above is to emphasize that the classical Lyapunov exponent $\tilde{\lambda}_{\langle \ln \rangle}$ is the average of logarithms. We can, however, compute also the logarithm of the average,
\begin{eqnarray}
\tilde{\lambda}_{\ln \langle . \rangle} &=& \lim_{t\rightarrow\infty} \frac{1}{t} \ln  \left[ \frac{1}{N_{ch}}  \sum_{k}^{N_{ch}} 
 \,e^{\lambda_k t}  \right]  \label{Eq:LogAve} \\
&=& \lambda_{max} + \lim_{t\rightarrow\infty} \frac{1}{t}\ln \left[  \frac{1}{N_{ch}} \sum_{k} 
 \,e^{(\lambda_k -\lambda_{max})t}  \right] 
\rightarrow \lambda_{max}. \nonumber
\end{eqnarray}
For $t\rightarrow \infty$, one might expect $\tilde{\lambda}_{\langle \ln \rangle}$ to converge to $\tilde{\lambda}_{\ln \langle . \rangle} $. But for finite times, as discussed in Ref.~\cite{Rozenbaum2017}, the quantum Lyapunov exponents $\Lambda_{\text{Q}}$ are closer to $\tilde{\lambda}_{\ln \langle . \rangle} $ than to $\tilde{\lambda}_{\langle \ln \rangle}$, because $\Lambda_{\text{Q}}$  is obtained from the logarithm of the fit. This closer proximity between $\Lambda_{\text{Q}}$ and $\lambda_{max}$ is confirmed for the Dicke model as well.

\begin{figure}[ht] 
\centering
\includegraphics[width=0.45\textwidth]{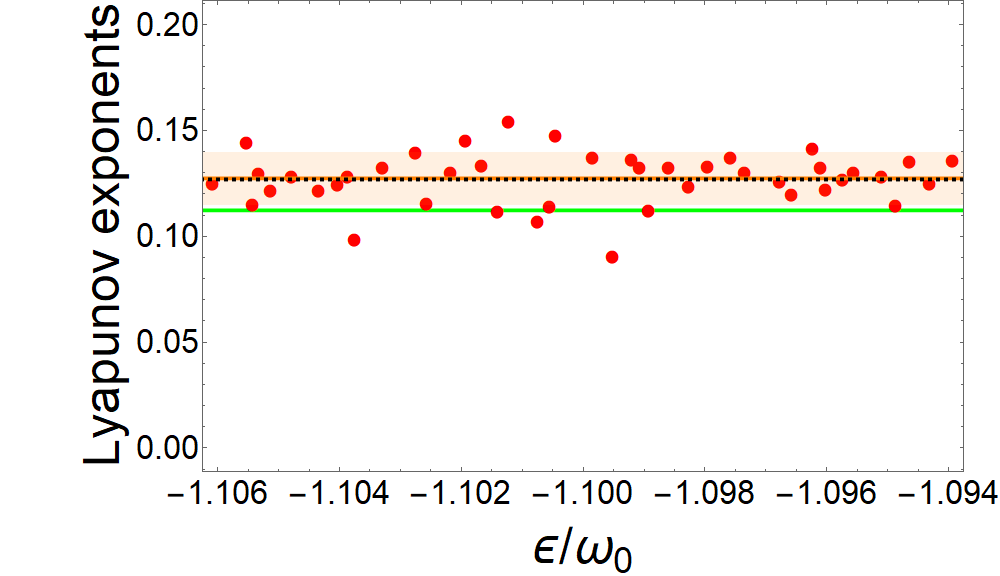} 
\caption{Comparison between the classical Lyapunov exponent $\tilde{\lambda}_{\langle \ln \rangle}$ (lower green horizontal line), the maximum classical Lyapunov exponent $\lambda_{max}$ (black dotted horizontal line), and the quantum Lyapunov exponents $\Lambda_{\text{Q}}$ (red circles) for $C_n^{qp}(t)$ for fifty-one states of different energies around $\epsilon/\omega_0\approx -1.1$. 
The solid orange line depicts the average value of $\Lambda_{\text{Q}}$ and the shaded region represents the standard deviation around it. 
}
\label{fig:OTOC_ener}
\end{figure}

In Fig.~\ref{fig:OTOC_ener}, we compare the classical Lyapunov exponent $\tilde{\lambda}_{\langle \ln \rangle}$ (lower green horizontal line), the maximum classical Lyapunov exponent $\lambda_{max}$ (black dotted line), the quantum Lyapunov exponents $\Lambda_{\text{Q}}$ (red circles) for the fifty-one energy states, and the average over the quantum Lyapunov exponents (orange solid line surperposed by the line for $\lambda_{max}$). The quantum exponents fluctuate due to the oscillations that modulate the exponential growth and to finite size effects; the standard deviation corresponds to the shaded area in the figure. Increasing the value of $j$ would reduce this uncertainty. While $\tilde{\lambda}_{\langle \ln \rangle} = 0.112$, the maximum classical Lyapunov exponent, $\lambda_{max} = 0.127$, coincides with the average value of the quantum Lyapunov exponent, $\bar{\Lambda}_{\text{Q}} = 0.126$, within its standard deviation $\sigma_\Lambda = 0.012$.

{\em Discussion.--}  We showed that for the Dicke model in the chaotic region, the OTOC grows exponentially fast in time with a rate comparable to the classical Lyapunov exponent. These results confirm that the quantum-classical correspondence established by means of the OTOC is not exclusive to one-body systems, but is valid also for interacting systems with more than one degree of freedom.  This work provides a proof-of-principle and should motivate similar studies in other interacting systems.

We stress that to clearly identify the quantum exponential growth and extract its rate, we need to have access to large system sizes. This was possible here, because we resorted to an efficient basis to construct the eigenstates.

The instrument of our analysis was the microcanonical OTOC [Eq.~(\ref{Eq:OTOC})] corresponding to the eigenstate expectation value of the commutator of two operators. Its use in stadium billiards~\cite{Hashimoto2017} prevented the observation of the quantum exponential growth, which was only possible with the introduction of Gaussian states~\cite{RozenbaumARXIV}. In our case, however, the eigenstates were excellent probe states for revealing the OTOC exponential growth. This is an important result for future studies of interacting systems, since the eigenstates are essential building blocks for thermal averages.

\begin{acknowledgments}
We thank L. Benet and T. Seligman for their useful comments. MABM, SLH, and JGH acknowledge J. Dukelsky for fruitful discussions in
the context of the Spanish project I-COOP2017:COOPB20289. PS is grateful to P. Cejnar for stimulating discussions. We acknowledge financial support from Mexican CONACyT project CB2015-01/255702, DGAPA- UNAM project IN109417 and RedTC. MABM is a post-doctoral fellow of CONACyT. PS is supported by the Charles University Research Center UNCE/SCI/013. LFS is supported by the NSF grant No. DMR-1603418.
\end{acknowledgments}

%%%%%%%%%%%%%%%%%%%% REFERENCES %%%%%%%%%%%%%%%%%%%%%
%\bibliography{biblioOTOC}

\begin{thebibliography}{70}
\bibitem [{\citenamefont {Scaffidi}\ and\ \citenamefont
  {Altman}()}]{ScaffidiARXIV}%
  \BibitemOpen
  \bibfield  {author} {\bibinfo {author} {\bibfnamefont {Thomas}\ \bibnamefont
  {Scaffidi}}\ and\ \bibinfo {author} {\bibfnamefont {Ehud}\ \bibnamefont
  {Altman}},\ }\href@noop {} {\enquote {\bibinfo {title} {Semiclassical theory
  of many-body quantum chaos and its bound},}\ }\bibinfo {note}
  {ArXiv:1711.04768}\BibitemShut {NoStop}%
\end{thebibliography}


%%%%%%%%%%%%%%%%%%%%%%%%%%%%%%%%%%%
\end{document}